\documentstyle[twocolumn,pra,aps]{revtex}

\begin{document}

\bibliographystyle{prsty}

\draft

\preprint{}

\tighten

\title{The Split Density Model: \\ A unified Description of 
Polarization and Array Dynamics for Vertical Cavity Surface Emitting Lasers}

\author{Holger F. Hofmann and Ortwin Hess}
\address{Institute of Technical Physics, DLR\\
Pfaffenwaldring 38--40, D--70569 Stuttgart, Germany}

\date{\today}

\maketitle

\begin{abstract}
We present a generalized rate equation model for the array and polarization
dynamics of general one or two dimensional arrays of Vertical Cavity Surface
Emitting Lasers (VCSELs). It is demonstrated that our model includes both 
the previous theory for edge emitting laser arrays and the theory of
polarization dynamics in quantum well VCSELs in a single unified description.
The model is based on the physical assumption of separated carrier densitiy 
pools individually coupled to different light field modes. 
These modes, defined by the separate gain profile of each pool, interact 
through the coherent dynamics of the light field derived from 
Maxwell's equations and represented
by the coefficients for index and loss guiding. 
The special case of two densities and two light field modes is solved 
and the
implications of the results for large VCSEL arrays are discussed.
Our analytic results show that typical solutions of the split density model  
range from phase locking to chaos, depending on the magnitude of  
the frequency coefficients. For weak coupling, the stable supermode is
always the mode of highest frequency. This indicates that 
anti-phase locking is the only stable phase locking possible in arrays.

\end{abstract}



\section{Introduction}

\label{sec:intro} 

If vertical cavity surface emitting lasers (VCSELs) are fabricated in the 
form of a densely packed two dimensional array as shown in Fig.\ref{Fig1}, 
their highly symmetric geometry around the axis of emission calls for an 
investigation of
both polarization dynamics and laser array interactions. Previously these
two aspects have been treated separately. Experimental investigations
of VCSEL arrays \cite{Ore92,War92,Hen96} usually cite theories derived for the one 
dimensional
case of side-by-side edge emitting laser arrays such as the theory introduced
by Winful and Wang in 1988 \cite{Win88}. On the other hand, a model for the 
description of polarization dynamics in individual VCSELs
was introduced by San Miguel, Feng and Moloney in 1995 \cite{Mig95}. 

Both models are rate equation models for type~B lasers. Adiabatic 
elimination of the dipole density is used to reduce the description
of the electronic system to carrier densities.
However, the similarity does not end here. In both models, the carrier 
density is split into separate pools, each of which couples to a different 
light field mode. The differences between the models arise only from the choice
of parameters which describe the dynamics of the electromagnetic field
and the diffusive exchange of carriers between the carrier density pools.

Since the physical assumptions of the models for polarization and array 
dynamics are so similar, we will show in the following that indeed a general 
model for the polarization and array dynamics of arbitrary VCSEL arrays can be 
formulated. The coefficients of this generalized split density model 
will then contain all the information necessary to describe any special case.
Therefore, the choice of the coefficient, which describe the physical
properties of the device studied, is critical and a good understanding
of the microscopic physics from which they can be derived is necessary to
distinguish between common and exotic cases.

In section~\ref{sec:phys}, the physical justifications for introducing 
split densities in both the polarization and the array dynamics is discussed.
Parameters for carrier diffusion and scattering are approximated.
In section~\ref{sec:form}, we present the general formalism of the split 
density model. The coefficients derived from Maxwell's equations are 
introduced and discussed.
In section~\ref{sec:two}, the special case of two densities and two light 
field modes is treated. Fixed points and limit cycles are analytically 
calculated.
In section~\ref{sec:general}, our analytic results of the two density model 
are applied to large arrays. A stability analysis is presented and 
conditions for chaotic dynamics are discussed.
In section~\ref{sec:concl}, we conclude by discussing the strengths and
weaknesses of the split density model as the basis of a description for
general semiconductor laser array dynamics.
         

\section{The physical justification of split densities}

\label{sec:phys}

\subsection{Polarization dynamics: separation of carrier density pools by 
angular momentum}

In quantum well VCSELs, the axis of symmetry of the quantum well 
structure is identical to the axis of laser light emission. Therefore,
the conservation of angular momentum may be applied to describe the emission 
process. 
Photons with an angular momentum of +1 (right circular polarized light) may
only be emitted by electron-hole pairs with a total angular momentum of +1 and
photons with an angular momentum of -1 (left circular polarized light) may
only be emitted by electron hole pairs with a total angular momentum of -1.
If only the bands closest to the fundamental gap are considered, electrons 
have an angular momentum of $\pm 1/2$ and holes have an angular momentum of 
$\pm 3/2$. Consequently, electrons with an angular momentum of +1/2 may only
recombine with holes of -3/2 angular momentum, emitting left circular 
polarized light. Electrons with an angular momentum of -1/2 may only
recombine with holes of +3/2 angular momentum, emitting right circular 
polarized light. 

Due to the conservation of angular momentum, the carrier density is 
therefore split into a carrier
density emitting only right circular polarized light and a carrier density
emitting only left circular polarized light.
This effect was first included in a rate equation model by San Miguel 
and coworkers in \cite{Mig95}. In that model, it is assumed that 
the two densities interact 
by spin flip scattering processes which give rise to a diffusive 
exchange of carriers between the two densities. 
Experimental investigations of quantum well structures indicate that this 
scattering rate is of the order of $10^{10} s^{-1}$, similar to the carrier
recombination rate \cite{Vin94}. The assumption of split densities should 
therefore
be well justified if the separation of bands in the quantum well structure is 
sufficient to exclude the participation of higher bands in the laser process.

\subsection{Array dynamics: spatial separation of carrier density pools}

In a typical semiconductor laser array, carriers are injected 
into spatially separated regions of the active medium. If the distance of 
separation is sufficiently large, gain guiding effects will give rise to 
spatially separated optical modes, each associated by the gain function
with its own carrier 
density pool. The localized modes are then coupled by Maxwell's
equations, since the evanescent parts of the neighbouring optical modes 
overlap.

The rate equation theory for laser arrays of this type was developed by
Winful and Wang in 1988 \cite{Win88}. However, in that early formulation,
 carrier diffusion is completely 
neglected. To check the justification of that assumption, we need to 
estimate the time it takes carriers to diffuse from one localized carrier
density pool to a neighbouring one. Noting that the microscopically
computed ambipolar diffusion constant of charge carriers in GaAs based lasers
is typically about $1 \mu m^2/ns = 10~cm^2/s$ \cite{Hes96a}, we can estimate
the diffusive carrier exchange rate between lasers separated by a few
micrometers to be close to $10^{10}s^{-1}$, similar to the spin 
relaxation rates 
expected for the polarization dynamics. However, the carrier
exchange rates depend on the square of the distance between the lasers, 
which implies that the split density model rapidly loses its validity as
the separation between elements of the array is reduced. In order to 
investigate the effects of diffusion, the possibility of carrier exchange 
between the densities is included in the generalized split density model
presented in~\ref{sec:form}.

In this model, carrier diffusion is described in terms of a carrier exchange 
rate.
Of course, this is just a crude approximation of the real physical process
of continuous diffusion which is quite different from the spin flip
scattering involved in the carrier exchange between carrier density pools
of different angular momentum. However, as long as the light field is only
sensitive to the average carrier density within one laser of the array
and higher transversal modes are neglected, the overall effect of diffusion
can be summarized in a single carrier exchange rate.

We cannot but note that some of the coupled mode theories frequently cited 
use the assumption of homogeonous and stationary gain, ignoring the
dynamical effects of spatially resolved carrier densities altogether
\cite{Ore92,Hen96,Kat84,But84,Agr85}. In type B lasers,
that assumption is only valid at the fixed points. As a consequence, those 
theories can not be used as starting points for an analysis of the stability 
and the dynamics of laser arrays.


\section{General formalism for the description of VCSEL arrays}

\label{sec:form}

\subsection{Rate equations for the carrier densities: 
injection, non-radiative recombination and diffusion}

The carrier densities of a VCSEL array can be separated into densities
$D_i$ associated with a position in the array and either left or right
circular polarization. In a two dimensional VCSEL array, $i$ represents
the coordinates $(r,c,\pm)$ for row number $r$, column number $c$ and plus
for right circular polarization, minus for left circular polarization.
When no electromagnetic field is present, the carrier dynamics 
including the effects of injection, non-radiative recombination and diffusion
are

\begin{equation}
\label{eq:density}
\frac{d}{dt}D_i = \mu_i - \gamma_i D_i - \sum_j T_{ij} (D_i-D_j),
\end{equation}

where $\mu_i$ is the effective local injection current, $\gamma_i$ the local 
non-radiative recombination rate and $T_{ij}=T_{ji}$ is the rate of 
diffusive carrier exchange between
the densities $D_i$ and $D_j$. 
Note that the model assumes transparency at $\mu = 0$. When comparing
the predictions of the model with experiment, it must be taken
into account that $\mu$ represents the additional injection current after
transparency has been reached. The real injection current is $\mu$ plus
a device dependent constant.
The non-radiative recombination rate
and the carrier diffusion can conveniently be combined into a single 
relaxation matrix of the densities by defining

\begin{equation}
\label{eq:relax}
\Gamma_{ij} = \delta_{ij}(\gamma_i + \sum_k T_{ik})  - T_{ij}.
\end{equation}

Equation \ref{eq:relax} describes
the diffusive relaxation to an equilibrium of injection and recombination
of electron-hole pairs. This constitutes the complete carrier dynamics at 
zero field. Typical relaxation rates are around $10^{10}s^{-1}$, 
as was discussed in~\ref{sec:phys}.

\subsection{Maxwell's equations of the electromagnetic field: gain, loss and coherent propagation}

The electromagnetic field inside the optical
cavity of the laser array can be described by the complex field amplitudes 
of localized modes $E_i$ of either left or right circular polarization.
Each mode is associated with one density pool. In a two dimensional array, $i$
represents the same coordinates $(r,c,\pm)$ given above. This is an essential
feature of the split density model. The split densities correspond to a
set of modes for the electromagnetic field which is different from
the Eigenmodes of Maxwell's equations. Instead, these modes are
Eigenmodes of the gain. They are the natural modes for the description of
relaxation oscillations and limit cycles as
well as the stability of fixed points. We will therefore chose this set
of Eigenmodes and not the supermodes to describe the laser array.
We note that we limit our theory to the fundamental mode of each laser in
the array and disregard possible effects of spatial hole-burning within each laser. 
A generalization to include higher modes is possible, but 
would make it more difficult to distinguish array effects from the internal
dynamics of individual lasers.

The dynamics of the electromagnetic field, including gain, loss and coherent
propagation are given by Maxwell's equations, which are linear wave equations.
In a general mode representation, they may be written as 

\begin{equation}
\label{eq:field}
\frac{d}{dt}E_i = w_iD_i(1-i\alpha)E_i 
                  - \sum_j (\kappa_{ij} + i\omega_{ij})E_j .
\end{equation}

The carrier density dependent gain is given by $w_iD_i$. The 
rate of spontaneous emission into the laser mode  $w_i$ is typically around
$10^6 s^{-1}$. Since most arrays will have very similar values for the
different $w_i$, the average and the deviations from this average may be
seperated into $w_i = \bar{w}(1+f_i)$. In this context, $f_i$ is the
measure of the relative fluctuation in the coupling between density pool
$i$ and mode $i$.  

The alpha factor $\alpha$ describes band structure effects which 
shift the gain profile to higher frequencies as the total density is
increased. Typical values range from 2 to 6. Effectively, this shift of
the gain profile introduces a carrier density dependent index of refraction.

$\kappa_{ij}$ describes the losses through the mirrors. In most cases, the
coefficients coupling different modes are very small and most of the losses
are given by $\bar{\kappa}\delta_{ij}$ with $\bar{\kappa}$ around $10^{12} 
s^{-1}$. Small irregularities and birefringence in the mirrors may cause 
non negligible loss guiding effects which can be represented by
small corrections $\bar{\kappa} l_{ij}$.

The matrix $\omega_{ij}$ describes the unitary coherent dynamics of Maxwell's 
equations based on the index of refraction of the array structure.
Neighbouring modes interact by their overlapping evanescent waves, while
right and left circular polarization modes are coupled by stress induced
birefringence commonly found in epitaxially grown semiconductor 
heterostructures \cite{Doo96}. Since the magnitude of these effects is 
highly dependent on
device architecture, it is difficult to estimate. However, their order of 
magnitude should be roughly between $10^{10} s^{-1}$ and $10^{12} s^{-1}$
to have a significant effect on the dynamics. For a given architecture,
numerical simulations may be used to determine the effective coupling 
constant. 

For stationary densities $D_i$, the dynamics of the electromagnetic field is
strictly linear. It is therefore possible to find orthogonal Eigenmodes 
of the dynamics including the cooperative effects of gain guiding ($w_iD_i$), 
loss guiding ($\kappa_{ij}$) and index guiding ($\omega_{ij}$). These modes
are the supermodes of the coupled mode theories. Note that Eigenmodes of 
different density distributions $D_i$ need not be orthogonal. Since 
semiconductor lasers are type B lasers, the dynamics of the $D_i$ cannot
be neglected and the modes defined by stationary $D_i$ are next to useless for 
the investigation of dynamical effects and stability analysis.

\subsection{Combined dynamics of the carrier density and the electromagnetic
field}

Now, the total dynamics of carrier density and electromagnetic field can be
constructed by adding only the term of induced emission, i.e. 
$-2w_iE_i^*E_iD_i$, to the carrier density dynamics. The resulting 
equations of motion for N split densities
and N complex field amplitudes representing, for example, an array of 
N/2 VCSELs with two polarizations each are
  
\begin{mathletters}
\begin{eqnarray}
\label{eq:general}
\frac{d}{dt}D_i &=& \mu_i - \sum_j \Gamma_{ij} D_j 
                  - 2\bar{w}(1+f_i)E_i^*E_iD_i \\
\frac{d}{dt}E_i &=& (\bar{w}(1+f_i)D_i(1-i\alpha) - \bar{\kappa})E_i
                  \nonumber \\ 
                & &  - \sum_j (\bar{\kappa}l_{ij} + i\omega_{ij})E_j .
\end{eqnarray}
\end{mathletters}
These two lines of equations already contain a large number of different
physical effects. The localized modes in the array are not 
only coupled by overlapping evanescent waves ($\omega_{ij}$) but also by
carrier diffusion ($\Gamma_{ij}$) and by variations in the losses ($l_{ij}$).
The injection current may be varied ($\mu_i$) and the coupling of the active
medium to the laser mode may also be different for each laser in the array
($f_i$).

In the following, we shall focus our attention on arrays pumped homogeneously
($\mu_i = \mu$) and with negligible variations in the gain and loss 
coefficients ($f_i = l_{ij} = 0$), resulting in 

\begin{mathletters}
\begin{eqnarray}
\label{eq:base}
\frac{d}{dt}D_i &=& \mu - \sum_j \Gamma_{ij} D_j 
                  - 2\bar{w}E_i^*E_iD_i  \\
\frac{d}{dt}E_i &=& (\bar{w}D_i(1-i\alpha) - \bar{\kappa})E_i  
                  - \sum_j i\omega_{ij}E_j .
\end{eqnarray}
\end{mathletters} 
This equation describes the counteractive effects of coherent field dynamics 
represented by $\omega_{ij}$ and carrier diffusion represented by
$\Gamma_{ij}$.

For the case of VCSEL polarization, the joint effects of coherent coupling
and non-negligible gain-loss anisotropy, $l_{ij} \neq 0$, has been 
investigated by the authors in a separate paper \cite{Hof97}. 


\section{The two density model}

\label{sec:two}

\subsection{Formulation of the rate equations}

To understand how the effects of coherent field dynamics and carrier 
diffusion influence the stability of phase locking and the transition
to chaos, we will now start our discussion with an investigation of the 
most simple case of two carrier density pools
and two light field modes. This case describes either the polarization
dynamics of a single VCSEL or an array of two edge emitting lasers. In these
contexts, the equations of the two density model have been studied previously 
by Martin-Regalado (polarization) \cite{Mar96} and by Winful and Wang (array)
\cite{Win88} on the basis of numerical simulations. In the following we will 
review the results of the two density model and analytically derive general
properties of array and polarization dynamics.

The equations for a symmetrical system of two carrier densitiy pools are 
defined by the
parameters $\bar{w}$, $\bar{\kappa}$, $\Gamma_{11}$, $\Gamma_{12}$,  and 
$\omega_{12}$. It is useful to keep in mind that $\Gamma_{11}$ is the sum
of the carrier diffusion rate $\Gamma_{12}$ and the non-radiative 
recombination rate $\gamma$. They read

\begin{mathletters}
\begin{eqnarray}
\label{eq:tdm}
\frac{d}{dt}D_1 &=& \mu - \Gamma_{11} D_1 + \Gamma_{12} D_2 
                  - 2\bar{w}E_1^*E_1D_1  \\
\frac{d}{dt}D_2 &=& \mu - \Gamma_{11} D_2 + \Gamma_{12} D_1 
                  - 2\bar{w}E_2^*E_2D_2  \\
\frac{d}{dt}E_1 &=& (\bar{w}D_1(1-i\alpha) - \bar{\kappa})E_1  
                  - i\omega_{12}E_2 \\
\frac{d}{dt}E_2 &=& (\bar{w}D_2(1-i\alpha) - \bar{\kappa})E_2  
                  - i\omega_{12}E_1 .
\end{eqnarray}
\end{mathletters} 
All rates are given by positive and real values. This is not a natural
assumption for $\omega_{12}$, since a phase transformation in one of the 
two modes would generally also change the phase factor of $\omega_{12}$. 
A real and positive  $\omega_{12}$ implies that the $E_1+E_2$ 
mode has a lower frequency than the $E_1-E_2$ mode for $D_1 = D_2$ (no gain
effects on the mode structure). In an array, this is a natural choice. 
If the modes describe orthogonal polarizations, however, the phase factor of
$\omega_{12}$ is an important definition of the phase relation between the 
modes.

\subsection{Stability analysis}

If the two subsystems were not coupled by $\omega_{12}$, they could be
described as two identical lasers operating at the same injection current.
Therefore, $D_1 = D_2$ and $E_1^*E_1 = E_2^*E_2$ is stationary. The 
introduction of $\omega_{12}$ removes the arbitrary choice of phase and
defines the supermodes with $E_1 = -E_2$ or $E_1 = E_2$ as the only possible
stationary solutions for the light field.

The exact stationary solutions are given by $D_1 = D_2 = \bar{\kappa}/
\bar{w}$ and $ E_1^*E_1 = E_2^*E_2 = \mu/2\bar{\kappa} - \gamma/2\bar{w} $.
To analyse the stability of the two possible supermodes, small fluctuations
around these stable values are defined:

\begin{mathletters}
\begin{eqnarray}
\label{eq:fluct}
\delta D &=& D_1 + D_2 - \frac{2\bar{\kappa}}{\bar{w}} \\
\delta n &=& E_1^*E_1 + E_2^*E_2 - \frac{\mu}{\bar{\kappa}}  
                                 - \frac{\gamma}{\bar{w}} \\
d        &=& D_1-D_2 \\
s_1      &=& E_1^*E_1 - E_2^*E_2 \\
s_2      &=& i(E_1^*E_2 - E_2^*E_1).
\end{eqnarray}
\end{mathletters}  

Note that in the case of polarization dynamics, $s_1$ and $s_2$ are
two of the three Stokes parameters. The third Stokes parameter is
defined as $ s_3 = E_1^*E_2 + E_2^*E_1 $ and is equal to plus or minus 
the total intensity for the two stationary solutions. 

Using these parameters, the linearized dynamics around the stationary point 
of the antisymmetric supermode with $E_1 = -E_2$ and $s_3 = -n_s$ is

\small
\vspace{1ex}
\begin{math}
\frac{d}{dt}\left(
\begin{array}{c}
\delta D\\ \delta n \\d\\s_1\\s_2
\end{array}\right) = 
\end{math}

\begin{equation}
\left(\! 
\begin{array}{ccccc}
-\!\gamma\! -\! \bar{w} n_s & - 2\bar{\kappa} & 0 & 0 & 0 \\
\bar{w}n_s & 0 & 0 & 0 & 0 \\
0 & 0 & -\!\gamma\! -\! \bar{w}n_s\! -\! 2 \Gamma_{12}  & -2\bar{\kappa} & 0\\
0 & 0 & \bar{w}n_s & 0 &  \omega_{12} \\
0 & 0 & \alpha \bar{w}n_s &  -\omega_{12} & 0
\end{array}\right)
\left(\!
\begin{array}{c}
\delta D\\ \delta n \\d\\s_1\\s_2
\end{array}\!\right), 
\\[2ex]
\end{equation}
\normalsize
where $n_s = E_1^*E_1 + E_2^*E_2$ is the total intensity of the stationary 
solution. To obtain the stability matrix for the 
symmetric supermode with $E_1 = E_2$ and $s_3 = +n_s$, one only has to 
change the sign of $\omega_{12}$.

The stability matrix can be separated into two blocks. One two by two
block describes the relaxation oscillations of the total intensity.
The remaining three by three block contains all the relevant information on
the interaction between the subsystems. For weak coupling, $s_1$ and $d$
show relaxation oscillations damped by diffusion, while $s_2$ will be
stabilized or destabilized exponentially by its interaction with $d$ via
$\alpha$.

The condition for the stability of a solution is, that the real parts
of all Eigenvectors of the matrix given above are negative. For the
symmetric supermode $E_1+E_2$, which is the lowest frequency mode, this
condition may be found by searching for a parameter set with one Eigenvalue
equal to 0. This is done by setting the determinant of the stability 
matrix to 0. The result of this analysis is 

\begin{equation}
\omega_{12} > 2\bar{\kappa}\alpha
              \frac{\bar{w}n_s}{\gamma + \bar{w}n_s + 2 \Gamma_{12}}.
\end{equation} 

Since we must assume that the diffusion rate $\Gamma_{12}$ is not much 
larger than the non-radiative recombination rate $\gamma$ 
for the split density model to be valid and since the rate of induced
emission $\bar{w}n_s$ cannot be
much larger than $\gamma$ for reasonable injection currents, this condition
can only be fulfilled if $\omega_{12}$ is as large as the loss rate 
$\bar{\kappa}$.
This would require frequency separations of THz between the supermodes,
which is unlikely if any reasonable gain, index or loss guiding exists to
separate the modes. Since such guiding effects are absolutely necessary to 
define the array, the low frequency solution is only stable if the description
of the device as an array of coupled lasers breaks down. It is therefore 
reasonable to assume that in a laser array guiding effects reduce 
$\omega_{12}$ to values below $\bar{\kappa}$. 
Consequently only the highest frequency supermode can be stable in a 
two mode array. If a different supermode is stabilized, this is most likely
not an array effect, but instead represents higher mode effects typical
of a broad area laser device \cite{Mer95,Hes96b}. 
In the case of polarization dynamics in VCSELs, a birefringence effect
of THz is unrealistically high. Experiments indicate that $\omega_{12}$
is more likely to be of the same order of magnitude as $\gamma$ \cite{Doo96}. 
Therefore, the stable polarization corresponds always to the one with
the highest frequency.
Of course, the stability condition for the low frequency mode can still
be fulfilled very close to threshold. This gives rise to a bistability 
which was reported in \cite{Mar96}. However, according to our estimates, 
this bistability should only occur extremely close to threshhold,
where noise effects dominate the laser dynamics.
This bistability is therefore not likely to be of any practical relevance.

The condition for stablility of the anti-symmetric supermode $E_1-E_2$
is found by searching for parameters permitting 
the existence of purely imaginary Eigenvectors. This indicates the point
at which the relaxation oscillations of $s_1$ and $d$ become undamped.  
The condition for damped relaxation oscillations is found to be

\begin{equation}
\alpha\omega_{12} < \gamma + \bar{w}n_s + 2 \Gamma_{12}.
\end{equation}  
This condition is always fulfilled if the damping effect of the diffusion 
rate $2 \Gamma_{12}$ exceeds the undamping effect of the coherent coupling 
$\alpha\omega_{12}$. Effectively, a moderate amount of diffusion 
stabilizes phase locking and prevents the transition to limit cycles and
chaos. If $\alpha\omega_{12}$ exceeds $\gamma + 2\Gamma_{12}$, a minimal 
intensity 
is necessary to stabilize phase locking. Possibly, this intensity is 
unrealistically high, so that injection currents far above threshold would
be necessary to stabilize the dynamics. Otherwise, a transition from 
stability to a limit cycle will be
observable as the injection current is lowered. 

\subsection{Dynamics close to the fixed point: damping and undamping
of relaxation oscillations}

The analytic results of the stability analysis show that both semiconductor 
laser arrays and birefringent VCSELs tend to emit light in the mode
of highest frequency. In order to provide a foundation for the discussion of 
limit cycles and for the extrapolation of the results to 
larger arrays, we will now take
a closer look at the dynamics of fluctuations around the fixed point.
This is done by finding the Eigenvectors and Eigenvalues of the stability
matrix. Although, in principle, approximations are not necessary to 
diagonalize the
two by two and three by three blocks of the stability matrix it helps the
interpretation of the physics if we make
use of the fact that for semiconductor lasers, the cavity loss rate 
$\bar{\kappa}$ is typically two orders of magnitude larger than the
non-radiative recombination rate $\gamma$. As a result, the fluctuations
of semiconductor lasers are dominated by fast relaxation oscillations of
the frequency $\nu = (2\bar{\kappa}\bar{w}n_s)^{1/2}$. The stability matrix
can then be diagonalized by treating the damping term $\gamma + \bar{w}n_s 
+ 2\Gamma_{12}$ and the coherent coupling $\omega_{12}$ as small perturbations
of the relaxation oscillations. In lowest order, the Eigenvalues $\lambda_i$ 
and the left and right Eigenvectors ${\bf a}_i$ and ${\bf b}_i$ are
\\[2ex]
\begin{mathletters}
\begin{math}
\lambda_{1/2} = -\frac{1}{2}(\gamma + \bar{w}n_s)  \pm i\nu 
\end{math}
\\[1ex]
\begin{center} 
\begin{math}
{\bf a}_{1/2}=\frac{1}{\sqrt{2}}
\left(
\begin{array}{ccccc}
\mp i \sqrt{\frac{\bar{w}n_s}{2\bar{\kappa}}} & 1 & 0 & 0 & 0 
\end{array}\right)
\end{math}
\end{center}

\begin{equation}
\label{eq:EV12}
{\bf b}_{1/2}=\frac{1}{\sqrt{2}}
\left(
\begin{array}{c}
\pm i \sqrt{\frac{2\bar{\kappa}}{\bar{w}n_s}} \\1\\0\\0\\0 
\end{array}\right)
\end{equation}
\\[2ex]
\begin{math}
\lambda_3 = -\alpha \omega_{12}  
\end{math}
\\[1ex]
\begin{center}
\begin{math}
{\bf a}_{3}=
\left(
\begin{array}{ccccc}
0&0&0&-\alpha&1 
\end{array}\right)
\end{math}
\end{center}

\begin{equation}
\label{eq:EV3}
{\bf b}_{3}= 
\left(
\begin{array}{c}
0\\0\\0\\0\\1 
\end{array}\right)
\end{equation}
\\[2ex]
\begin{math}
\lambda_{4/5}=-\frac{1}{2}(\gamma + \bar{w}n_s + 2\Gamma_{12} 
                           - \alpha\omega_{12}) \pm i \nu   
\end{math}
\\[1ex]
\begin{center}
\begin{math}
{\bf a}_{4/5}=\frac{1}{\sqrt{2}}
\left(
\begin{array}{ccccc}
0&0&\mp i \sqrt{\frac{\bar{w}n_s}{2\bar{\kappa}}} &1&0 
\end{array}\right)
\end{math}
\end{center}

\begin{equation}
\label{eq:EV45}
{\bf b}_{4/5}=\frac{1}{\sqrt{2}}\left(
\begin{array}{c}
0\\0\\ \pm i \sqrt{\frac{2\bar{\kappa}}{\bar{w}n_s}}    \\ 1 \\ \alpha 
\end{array}\right).
\end{equation}
\end{mathletters}

In (\ref{eq:EV12}) and (\ref{eq:EV45}), solutions 1 and 4 refer to the 
upper sign and solutions 2 and 5 refer to the lower sign, respectively.

The solution 1/2 describes the linearized relaxation oscillations of the 
total intensity and carrier density, solution 3 describes the stabilization of 
anti-phase locking and solution 4/5 describes the relaxation oscillations
of the difference in the intensities and carrier densities. 

The solutions 3 and 4/5 describe the supermode stability we are interested in.
We note that an important feature of the Eigenvectors is their 
non-orthogonality. This
is caused by the $\alpha$ factor. The $\alpha$ factor converts density
fluctuations into phase fluctuations. An example to illustrate this effect
can be given by the relaxation dynamics of a fluctuation in the intensity
difference $s_1$,
described at time $t=0$ by the vector $(0,0,0,1,0)$. As shown in 
Fig.~\ref{Fig2}, 
this intensity fluctuation induces a phase fluctuation proportional to
$\alpha$ as it relaxes.
 
The relaxation rates of the solution are influenced by this correlation
between phase fluctuations and intensity fluctuations. 
In solution 3, the $\alpha$ factor converts the time 
derivative of the intensity difference, $d \! / \! dt \; s_1 
= \omega_{12}s_2$, into a time derivative of the phase, 
$\lambda_3 s_2 = \alpha \; d \! / \! dt \; s_1$, causing
stability or instability, depending on the mode.
In solution 4/5, the time derivative of the phase is likewise converted
into a time derivative of the intensity difference, increasing or
decreasing stability. If the phase difference is stabilized, the
relaxation oscillations are destabilized and vice versa. Mathematically,
this is necessary because the sum of all Eigenvalues $\lambda_i$ must
be equal to the trace of the stability matrix.

Since the $\alpha$ factor increases frequency with increasing carrier
densities, the equations show that only the supermode with highest
frequency is stabilized in the presence of relaxation oscillations. 
However, the stability analysis has shown that strong coupling may 
destabilize this
solution as well. In the Eigenvalues of solution 4/5, this is expressed 
by the negative contribution of $\alpha\omega_{12}$. 

To understand the undamping effect given by $\alpha\omega_{12}$, let us recall
that $s_1$ and $s_2$ oscillate in phase, with $s_2 = \alpha s_1$.
Consequently, the matrix element $\omega_{12}$
causes an undamping of this oscillation by adding a term equivalent to 
$+\alpha\omega_{12}s_1$ to the time derivative of $s_1$. Effectively, we thus
encounter the surprising property that the
coherent dynamics of the light field pumps intensity from the mode at low 
intensity to the mode at high intensity. This destabilizing effect is 
countered by the carrier density relaxation effects when the carrier
density difference $d$ is maximal and the light field intensity difference
$s_1$ is zero. In this way, non-radiative ($\gamma$) and radiative 
($\bar{w}n_s$) carrier recombination and diffusion ($2\Gamma_{12}$)
stabilize the relaxation oscillations by damping $d$, while coherent coupling 
($\alpha\omega_{12}$) destabilize them by undamping $s_1$.

\subsection{Low amplitude limit cycle}

If the relaxation oscillations are undamped, the result is a low amplitude 
limit cycle. For small amplitudes, this limit cycle still resembles
relaxation oscillations with   

\begin{mathletters}
\begin{eqnarray}
d(t)   &=& A_0 \sqrt{\frac{2\bar{\kappa}}{\bar{w}n_s}} cos(\nu t) \\
s_1(t) &=& A_0 sin(\nu t).
\end{eqnarray}
\end{mathletters}   

The damping term is still linear, since the diffusion and the carrier
recombination terms are linear over a wide range of carrier densities.
The undamping term is more complicated. It results from the phase dynamics

\begin{equation}
tan(\phi) = i\frac{E_1^*E_2-E_2^*E_1}{E_1^*E_2+E_2^*E_1} = \frac{s_2}{s_3}
\end{equation}
of the phase difference $\phi$ between $E_1$ and $E_2$.
For small fluctuations, $s_3 \approx -n_s$ and $tan(\phi) \approx \phi$,
so that $s_2 \approx - n_s \phi$. If non-linear terms are included, the 
phase dynamics of the limit cycle is given by

\begin{equation}
\phi (t)   = -\frac{\alpha}{n_s} A_0 sin(\nu t).
\end{equation}

Since the total intensity $n_s$, the 1st Stokes parameter $s_1$
and the ratio between the 2nd and 3rd Stokes parameters (i.e. $tan(\phi)$)
are known,
we can determine $s_2$ using $n_s^2 = s_1^2 + s_2^2 +s_3^2$:

\begin{equation}
s_2 (t)   = \sqrt{n_s^2 - s_1(t)^2} sin(\phi(t)).
\end{equation} 
 
The 3rd order correction to the linear dynamics of $s_2$ is then given
by

\begin{equation}
s_2 (t)   = \alpha s_1(t) (1 - s_1(t)^2 \frac{3+\alpha^2}{6n_s^2}+\ldots).
\end{equation} 

For small $s_1/n_s$, the term of 3rd order in $s_1$ reduces and flattens 
the maxima of $s_2$. To determine the time averaged effect on the amplitude 
$A_0$, the contributions to the time derivative of $s_1$ in phase with
the oscillations must be found. This is the Fourier component of $sin(\nu t)$.
Since the Fourier component of
$sin(\nu t)$ in $sin^3(\nu t)$ is $3/4$, the total time averaged damping is

\begin{equation}
\frac{1}{2}(\gamma + \bar{w}n_s + 2\Gamma_{12} - \alpha \omega_{12}) 
+ \frac{3}{8}\alpha\omega_{12}\frac{A_0^2}{n_s^2}\frac{3+\alpha^2}{6}.
\end{equation}

Setting this term to 0, the amplitude of the limit cycle is found to be

\begin{equation}
A_0 = 2 n_s \sqrt{\frac{2}{3+\alpha^2}(1-\frac{\gamma+\bar{w}n_s+2\Gamma_{12}}
{\alpha\omega_{12}})}.
\end{equation}  

This equation describes the transition from stability to a limit cycle 
possible for strong coupling and low diffusion. The intensity dependent
contribution to the stability is the rate of induced emission, $\bar{w}n_s$. 

The amplitude of the limit cycle which has been calculated numerically in 
the paper on
laser arrays by Winful and Wang \cite{Win88} can now be determined analytically
using this formula. That limit cycle was calculated at an injection current
of 1.1 times threshold, while stability is reached at 2 times 
threshold for the parameters used. Fig.\ref{Fig3} shows the dependence of 
$A_0$ on the injection rate for this choice of parameters.


\section{Generalization of the two density model results to larger arrays}

\label{sec:general}

\subsection{Effective next neighbour interaction}

Our analysis shows that in the case of stability,
two interacting lasers will emit light 180 degree out of phase. This 
result can also be applied to larger arrays. Usually, the coupling constants
$\omega_{ij}$ only couple next neighbours. If all $\omega_{ij}$ are real
and positive, the stable mode will be the one in which next neighbours 
emit light 180 degree out of phase.
Note that in the case of a VCSEL array, the two circular polarizations must be 
considered as separate members of the array, coupled by birefringence.
In this sense, VCSEL polarization adds a third dimension to the otherwise
two dimensional array.

In most conventional geometries, it is very easy to find a single stable
supermode in which all next neighbours emit exactly out of phase. For
example, a square array of VCSELs will emit light of a single linear 
polarization in the highest frequency supermode. In the far field, this
corresponds to four intensity maxima emitted into directions tilted
towards the corners of the array. This result has been observed 
experimentally \cite{Ore92}, providing evidence that the split density 
model is a valid description for realistic devices.

In unconventional geometries such as triangular arrays or spatially varying
birefringence, there may be more than one stable supermode. The stationary
solutions may be obtained by letting the field amplitudes $E_i$ be equal to 
$+E_0$ or $-E_0$.
Among these modes, promising candidates for stability are always the 
arrangements with the
highest possible number of phase changes between next neighbours. 

Once the stable modes have been identified using the rule that next neighbours
try to achieve anti-phase locking, the stability of this mode with regard to
phase fluctuations and relaxation oscillations may be investigated. 
Since we can assume that the fastest time scale is that of the relaxation
oscillations, the two effects may be separated. The justification for this 
separation is the same as the one underlying the stability analysis of 
\ref{sec:two}: The Eigenvectors and
Eigenvalues of the stability matrix may be determined by treating $\omega_{ij}$
and the effects of damping and diffusion as small perturbations of the fast
relaxation oscillations. The 3N dimensional system of equations for the
carrier densities and the real and imaginary parts of the fields of N lasers
can then be separated into a system with 2N dimensions describing relaxation 
oscillations and one of N dimensions describing phase coupling.

In the following, we limit the coupling terms to next neighbour interactions, 
allowing a complete
stability analysis of large symmetric laser arrays.  

\subsection{Stability of the anti-phase locked supermode}

First, we will investigate the N dimensional stability matrix which 
describes the phase coupling of the array. In the two density model,
the phase fluctuation was described by $s_2 \approx -n_s \phi$, where the
phase difference $\phi = \delta \phi_1 - \delta \phi_2$ represents a measure 
of the difference between the phase fluctuations in system 1 and system 2.
If $\delta\phi_i$ denotes the phase fluctuation in system i, the two density 
stability matrix for phase fluctuations is

\begin{equation}
\label{eq:pfluc}
\frac{d}{dt}\left(
\begin{array}{c}
\delta \phi_1 \\ \delta \phi_2 
\end{array}\right) = \left(\! 
\begin{array}{cc}
-\frac{\alpha\omega_{12}}{2}& + \frac{\alpha\omega_{12}}{2}\\
+\frac{\alpha\omega_{12}}{2}& - \frac{\alpha\omega_{12}}{2}
\end{array}\right)
\left(
\begin{array}{c}
\delta \phi_1 \\ \delta \phi_2
\end{array}\right). 
\\[1ex]
\end{equation}

This result may be generalized to larger arrays by using (\ref{eq:pfluc})
as next neighbour interaction of phase fluctuations in
arbitrary arrays. The general equation for the dynamics of phase
fluctuations is then given by

\begin{equation}
\frac{d}{dt}\delta\phi_i = \sum_j \frac{\alpha\omega_{ij}}{2}
                                  (\delta\phi_j - \delta\phi_i).
\end{equation}

This equation describes the linear relaxation of phase fluctuations 
in an array of arbitrary size and geometry. The stability matrix is
defined by the $\alpha\omega_{ij}$. Note, that the relaxation has the 
properties
of a diffusion such as the one described by $T_{ij}$ for the carrier 
densities as introduced in \ref{sec:form}. This indicates
that local phase changes diffuse to the neighbouring systems until
the whole array is again phase locked. For an array with lattice
constant $a_0$, the phase diffusion constant is approximately given by
$a_0^2\alpha\omega_{ij}$. Thus, for micrometer scale arrays, typical phase 
diffusion constants are probably around $1 \mu m^2/ns$. This is the same 
order of magnitude as the carrier diffusion constant, even though the physical 
properties defining each are quite independent of each other. 

\subsection{Anti-diffusion and the route to chaos}

The remaining 2N variables of the stability matrix describe 
relaxation oscillations of the carrier densities and the light field 
intensities of the array.

In the two density model, the relaxation oscillation amplitude of both the 
total intensity and the intensity difference 
are damped by the non-radiative and the radiative recombination of carriers,
i.e. $\gamma + \bar{w}n_s$. In addition,
the oscillations of the intensity difference are damped by diffusion,
$2\Gamma_{12}$ and undamped by the coherent interaction between the two 
systems, $\alpha\omega_{12}$. 
Again the coherent interaction acts as a diffusion process. However,
this time the diffusion constant is negative: the coherent coupling gives 
rise to  anti-diffusion, increasing the intensity difference between next 
neighbours.

If we denote the local amplitude of the relaxation oscillations in the 
subsystem $i$ by the complex value $A_i$, the two density result may be 
generalized to arbitrary arrays on the basis of the same principles
applied in the previous subsection to the phase fluctuations. The dynamics 
of the $A_i$ averaged
over several relaxation oscillations is then given by 
 
\begin{equation}
\frac{d}{dt}A_i = - \frac{1}{2} 
                   ((\gamma_i + \bar{w} n_s) A_i 
                  + \sum_j (T_{ij}-\frac{\alpha\omega_{ij}}{2})
                           (A_i - A_j)).
\end{equation}
Note, that $\Gamma_{ij}$ has been decomposed into the contributions from
non-radiative carrier recombination $\gamma_i$ and from diffusion $T_{ij}$,
because they contribute to different types of effects. 
$\gamma_i + \bar{w}n_s$ causes an overall relaxation of the oscillations, 
while the
diffusion and anti-diffusion terms $T_{ij}-\alpha\omega_{ij}/2$ are sensitive 
only to amplitude 
differences between nearest neighbours. For $\gamma_i = \gamma_0 = constant$,
the relaxation effect stabilizes all forms of relaxation oscillations 
equally well. The diffusion stabilizes especially the relaxation
oscillations with an amplitude which varies strongly between neighbouring 
systems.
Consequently, these relaxation
oscillations are also undamped most effectively when 
$T_{ij}<\alpha\omega_{ij}/2$. In a regular lattice, the 
least stable relaxation oscillation in the presence of such anti-diffusion 
is the one in which $A_i = - A_j$ for
next neighbours. In a square lattice with constant coherent interaction
$\omega_{nn}$ and constant diffusion $T_{nn}$ between next neighbours,
the stability condition for this oscillation is

\begin{equation} 
\alpha\omega_{ij} < 2 T_{ij} + \frac{1}{4}(\gamma_i A_i + \bar{w}n_s) 
\end{equation}

When $\alpha\omega_{ij}$ exceeds this limit, the relaxation oscillations 
become undamped and
the near field intensity oscillates in a chessboard pattern. Since this 
implies that the intensities of all VCSELs emitting in the same phase 
oscillate in parallel, this oscillation should show up as a small intensity 
peak in the otherwise dark center of the far field pattern. 

For stronger anti-diffusion, more and more relaxation oscillations become 
undamped and the amplitude of the oscillations increases. The chessboard 
pattern of the oscillations will then be washed out and non-linear effects 
shift the
frequencies and add higher harmonics. For very strong coupling, the 
neighbouring VCSELs will switch on and off in an unpredictable chaotic
fashion. 


\section{Practical implications and open questions}

\label{sec:concl}

The split density model is the most simple way of describing the
array and polarization dynamics of VCSELs. It is also applicable to
the array dynamics of conventional edge emitting arrays. 

For the most interesting case of large symmetric arrays, the results
of the split density model are strikingly simple and clear: if the
separation between the lasers in the array does not break down because
of rapid diffusion, next neighbours lock into anti-phase laser emission.
Consequently, only the highest frequency supermode is stabilized.

For VCSEL arrays, this behavior has indeed been observed experimentally 
in \cite{Ore92,War92,Hen96}.
In \cite{Ore92}, this effect is explained as the result of higher losses 
between the VCSELs in the array. In the split density model, this 
corresponds to a coupling via 
the $l_{ij}$ terms introduced in~\ref{sec:form}. Such terms do indeed
stabilize the highest order supermode to a certain degree. However,
since the $l_{ij}$ are
proportional to the weak intensities, while the $\omega_{ij}$ terms
are proportional to the fields between the VCSELs
, this effect is likely to be much smaller than the coherent interaction. 
We therefore conclude that the explanation given in \cite{Ore92} for the 
experimentally observed anti-phase locking may be a misinterpretation 
caused by the neglect of carrier dynamics in many coupled mode theories
\cite{Kat84,But84,Agr85}.
Those theories certainly ignore the stabilizing effects of carrier dynamics
and coherent interaction, which cannot be neglected in type B lasers.

Another aspect of the split density model, pioneered in the numerical 
work by 
Winful and Wang \cite{Win88}, is the existence of limit cycles and chaos
in semiconductor laser arrays. 
We have calculated the stability limit including carrier diffusion effects 
and have shown that
stability is not as unlikely as predicted in the original paper by
Winful and Wang. Indeed, carrier diffusion may stabilize phase locking. 

A careful analysis of the fluctuations in the center of a far field pattern 
such as the one observed in \cite{Ore92} could reveal whether
the array is well stabilized by diffusion (weak fluctuations in the center),
shows anti-diffusion (fluctuations have a maximum in the center) or is already
oscillating in a weak limit cycle (coherent light is emitted into the center
of the far field). Possibly, the correct range of parameters necessary for
limit cycles and weak chaos could be realized experimentally by varying the
distance between the VCSELs to change diffusion effects and by using different
types of optical guiding. For larger separations, pure gain guiding may be
sufficient, while loss guiding is better for very small arrays.
    
In the same spirit, VCSEL polarization may be investigated, using
the quantum well width to change scattering rates and stress to induce
birefringence. Here, too, the frequency spectrum of the polarization
orthogonal to the laser mode reveals a lot about the mechanism of 
stabilization in the VCSEL and may actually include coherent emissions
from a limit cycle. Possibly, some of the VCSELs classified as having
only poor polarization stability in reality show a well stabilized high
frequency limit cycle and it is 
the time average of this limit cycle which is observed in polarization 
measurements.

Of course, when interpreting experimental results, the underlying assumptions 
made in the split density model must be kept in
mind. If the diffusion or the spin flip scattering is too strong, the split
densities must again be considered as a single carrier density and the 
supermodes
become higher modes of a single laser with many modes. Consequently, the 
split density 
model can be used to describe the transition from split to unified density.
This has been attempted for polarization dynamics in quantum well VCSELs
by Martin-Regalado in \cite{Mar96}. However, that
approach is only valid for a narrow range of parameters and it may be 
debatable if  the limitation to the lowest quantum well bands implied in
\cite{Mig95,Mar96} is still 
justified in quantum wells with such a strong spin flip scattering.

If the injection current is increased, higher order modes of the individual
lasers in the array may also participate in the dynamics. However, to 
simplify the presentation we have disregarded multi-transverse modes of an
individual laser. When considering that case, one has to assume that 
the dynamics is not governed by array effects, but instead by phenomena 
typical of broad area laser dynamics. The transition between the two 
regimes can be seen in numerical
simulations of gain guided edge emitting arrays \cite{Mer95,Hes96b}. 
In order to simulate the 
effect in the framework of the split density model, higher modes must be
assigned to each density. However, this rapidly reduces the simplicity
which makes the model so attractive.

In conclusion, we have shown that the split density model provides a 
simple description of semiconductor laser arrays and is equally 
useful for the description of VCSEL polarization. The results clearly
show that the mode with the highest frequency is stabilized. This is in
agreement with the experimental results reported in \cite{Ore92,War92,Hen96} 
and indicates that the handwaving explanation of the anti-phase locking
given in \cite{Ore92} may indeed be wrong.
We have seen that strong coupling may undamp the relaxation oscillations, 
leading through limit cycles to chaos. However, carrier difusion and both 
radiative and
non-radiative carrier recombinations stabilize the anti-phase locked
mode, which makes the stable operation of a VCSEL array more likely 
than the chaotic case.



%
%

%

\begin{figure}
\caption{Schematic geometry of a VCSEL array.}
\label{Fig1}  
\end{figure}

\begin{figure}
\caption{Relaxation dynamics of the phase variable $s_2$ and the intensity 
difference $s_1$ induced by a fluctuation in the intensity difference $s_1$ 
at time $t=0$. The choice of parameters is fully characterized by $\alpha = 3$,
$\alpha\omega_{12}=\nu/15$ and $(\gamma+\bar{w}n_s+2\Gamma_{12} 
- \alpha\omega_{12})/2 = \nu/5$. This example illustrates clearly, that
fluctuations in the intensity difference $s_1$ induce strong fluctuations in
the phase difference described by $s_2$.}
\label{Fig2}  
\end{figure}

\begin{figure}
\caption{Dependence of the limit cycle amplitude $A_0$ on the carrier 
injection rate. The choice of parameters corresponds to that of [4] 
: $\alpha = 5$, $\Gamma_{12}=0$ and $\alpha\omega_{12}=2 \gamma$. 
The injection
rate in units of the threshold injection rate is equal to 
$1+\bar{w}n_s/\gamma$. For this choice of parameters, the transition from
the limit cycle to stabilization by induced emission is within the 
experimentally accessible range of injection current and would therefore
be observable.}
\label{Fig3} 
\end{figure}

%

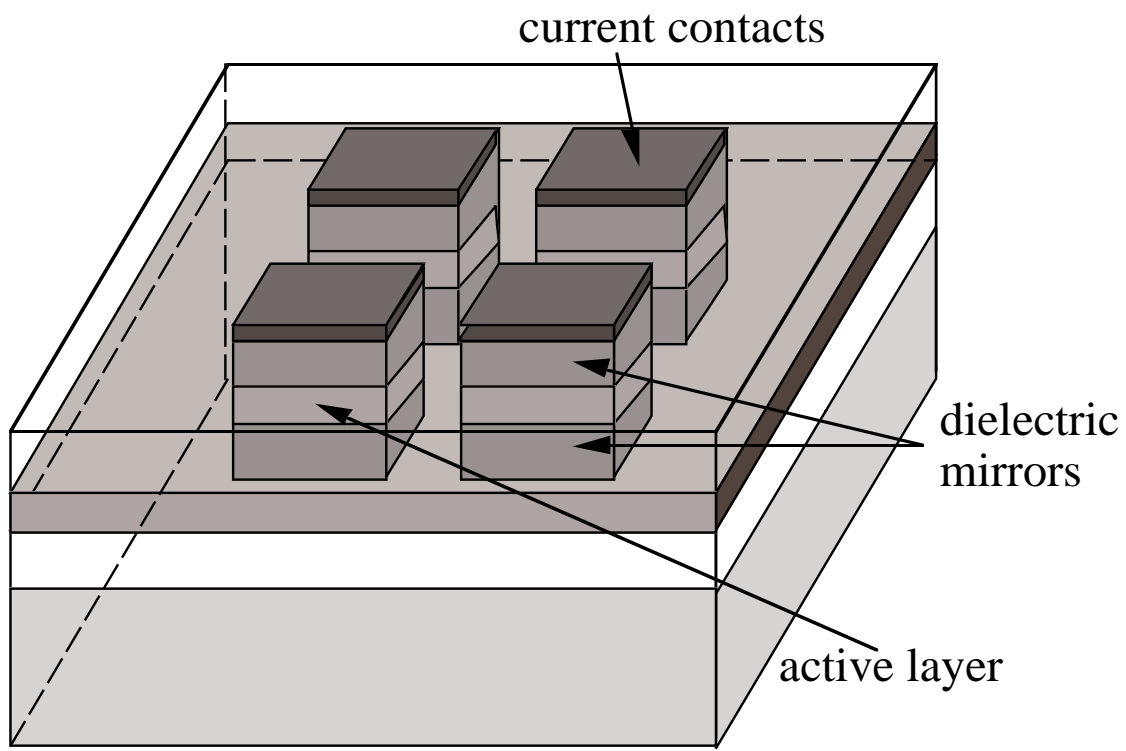
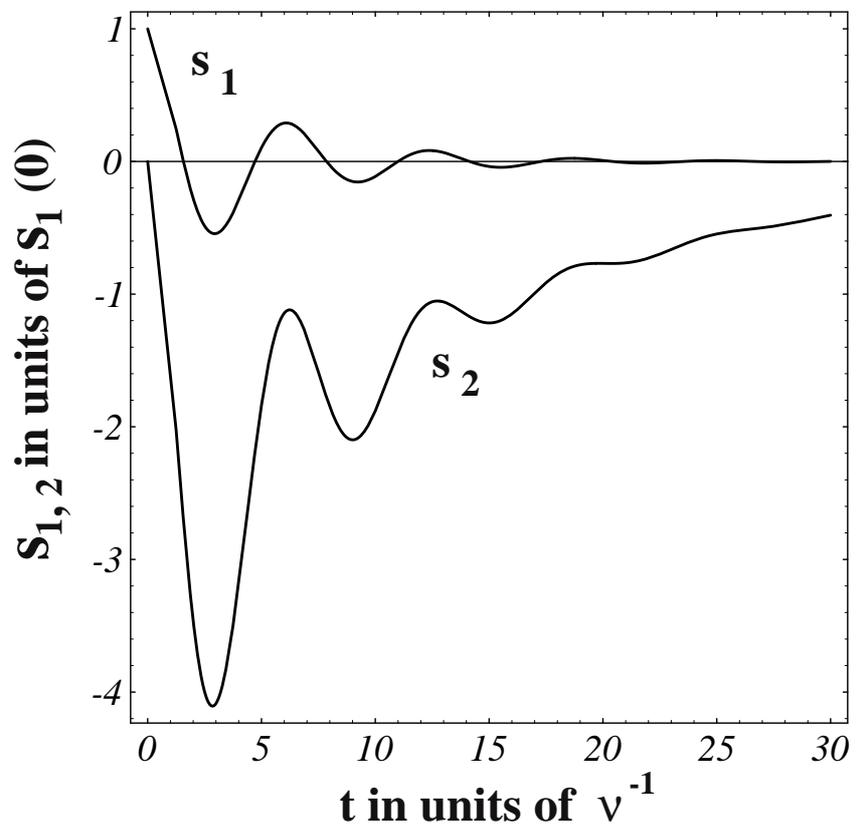
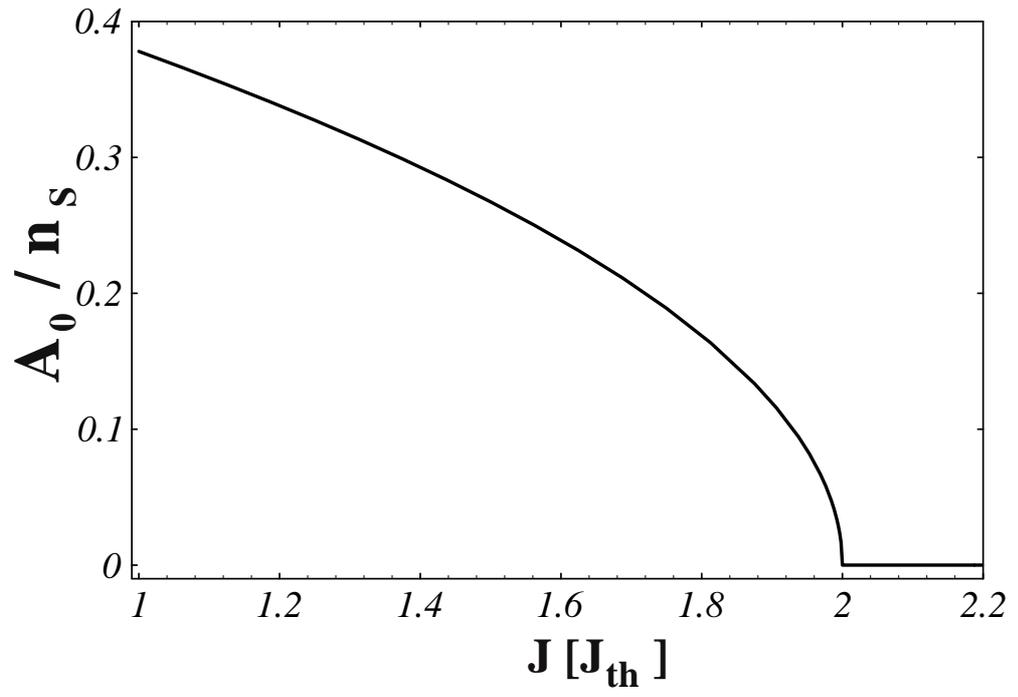
\end{document}